\documentclass[twocolumn,amssymb,fleqn, twocolumns]{revtex4} 
\usepackage{epsfig,amssymb,amsmath,graphicx,subfigure,hyperref}  
\usepackage{color}

\usepackage{multirow} 
\usepackage{tabularx}

\usepackage{diagbox}

\newcommand{\blue}[1] {{\color{blue} #1}}

\newcommand{\be} {\begin{eqnarray}  }
\newcommand{\ee} {\end{eqnarray} }

\newcommand{\f} {\frac }
\newcommand{\la} {\langle }
\newcommand{\ra} {\rangle }

\newcommand{\lb} {\left( }
\newcommand{\rb} {\right) }

\begin{document}

\title{Defect-free and defective adaptations of crystalline sheets to stretching deformation}
\author{Ranzhi Sun and Zhenwei Yao}
\email[]{zyao@sjtu.edu.cn}
\affiliation{School of Physics and Astronomy, and Institute of Natural Sciences,
	Shanghai Jiao Tong University, Shanghai 200240, China}
\begin{abstract} 
  The elastic response of the crystalline sheet to the stretching deformation in
  the form of wrinkles has been extensively investigated. In this work, we
  extend this fundamental scientific question to the plastic regime by exploring
  the adaptations of crystalline sheets to the large uniaxial mechanical
  stretching. We reveal the intermittent plastic shear deformations leading to
  the complete fracture of the sheets wrapping the cylinder.  Specifically,
  systematic investigations of crystalline sheets of varying geometry show
  that the fracture processes can be classified into defect-free and
  defective categories depending on the emergence of topological defects. We
  highlight the characteristic mechanical and geometric patterns in response to
  the large stretching deformation, including the shear-driven intermittent
  lattice tilting, the vortex structure in the displacement field, and the
  emergence of mobile and anchored dislocations as plastic excitations. The
  effects of noise and initial lattice orientation on the plastic deformation of
  the stretched crystalline sheet are also discussed.  These results advance our
  understanding of the atomic level on the irreversible plastic instabilities of
  2D crystals under large uniaxial stretching and may have potential practical
  implications in the precise engineering of structural instabilities in
  packings of covalently bonded particulate systems.
\end{abstract}

\maketitle

\section{Introduction}

Stretching a free-standing thin elastic sheet leads to the nonintuitive
wrinkling behavior for relaxing the in-plane strain incompatibility generated by
the Poisson effect~\cite{cerda2003geometry,Marder2007,audoly2010elasticity}. The
wrinkles are parallel to the direction of the applied tension in both
rectangular~\cite{cerda2003geometry,chopin2014roadmap} and annular elastic
sheets~\cite{yu1986elastic,geminard2004wrinkle}, and they provide a means for
mechanical characterization of stretchable soft solid membranes that have
practical
implications~\cite{harris1980silicone,cerda2002wrinkling,lacour2003stretchable,wang2022mechanics}.
The wrinkle structure arising in various stretched elastic systems has been
extensively studied by
experimental~\cite{geminard2004wrinkle,sharon2007geometrically,li2012mechanics},
computational~\cite{friedl2000buckling,lecieux2012numerical,wang2013physics}, and
theoretical~\cite{yu1986elastic,cerda2003geometry,puntel2011wrinkling,Grason2013,chopin2014roadmap}
approaches.  While the elastic response of the elastic sheet to the stretching
deformation in the form of wrinkles has been systematically investigated,
irreversible plastic instabilities of an intact crystalline sheet, which are inevitable in the large stretching
regime of real systems~\cite{sharon2007geometrically,chopin2016disclinations,jules2020plasticity},
have not yet been fully explored especially on the microscopic level. 
Understanding the plasticity of an intact crystalline sheet fabricated by
regular packings of particles in triangular lattice under large stretching
deformation is of fundamental and practical significance. Especially, it has
a strong connection to the important subject of mechanical instabilities involved
in a host of physical processes, including the failure of 2D crystalline
materials~\cite{miguel2011laminar,negri2015deformation,Mitchell2016,liu2022recent,chen2022geometry,sun2024plastic},
crystal growth on
interfaces~\cite{politi2000instabilities,meng2014elastic,kohler2015relaxation,ma2019growth},
and 2D assemblies of
colloids~\cite{dinsmore2002colloidosomes,Palacci2013b,bowick2016colloidal} and
viral shells~\cite{caspar1962physical,lidmar2003virus,roos2010physical}.

The goal of this work is to explore how the crystalline sheet consisting of
Lennard-Jones (L-J) particles in triangular lattice adapts to the large uniaxial
mechanical stretching, focusing on the characteristic mechanical and geometric
structures arising from the irreversible plastic instabilities. In our model, to
avoid any complication caused by the boundary condition, the stretching
deformation is imposed on the crystalline sheet by wrapping it around the
cylindrical substrate that is subject to controllable gradual expansion; the
cylindrical surface provides the geometric constraint only and no friction is
involved. The formally simple L-J potential is employed to study the plastic
deformation process for its featured energy minimum structure, and it has also
been extensively used to model various chemical and physical bonds in condensed
matter systems~\cite{jones1924determination,israelachvili2011intermolecular}.
This model contains the essential elements for us to study plastic instabilities
of mechanically stressed regular packings of particles that can be modeled as an
isotropic elastic sheet in the continuum
limit~\cite{Landau1986,audoly2010elasticity}. Note that the model of the
crystalline sheet wrapping on the cylinder is also connected to the
cylindrical crystal, which arises in many contexts, including the cylindrical
packings of disks and spheres, phyllotaxis (``leaf arrangement") in botany,
colloids on the surface of a liquid film coating a solid cylinder, and cell
walls of rod-shaped
bacteria such as \textit{E. coli}~\cite{mughal2012dense,amir2012dislocation,nelson2012biophysical,amir2013theory,mughal2014theory,beller2016plastic}.


The main results of this work are presented below. We first perform
preliminary theoretical analysis on the homogeneous elastic deformation of the
crystalline sheet under small stretching and confirm the reliability of the
computational approach. In the regime of large stretching, we reveal the
intermittent nature of the plastic shear deformations leading to the complete
fracture (disconnection) of the crystalline sheet wrapping the cylinder.  The
fracture processes are classified into defect-free and defective categories
depending on the emergence of topological defects. In defect-free plastic shear
deformations, we observe the tilting of the entire lattice, which is
quantitatively analyzed by a geometric model. We also highlight the vortex
structure formed in the shear-driven displacement field and the associated glide
motion of dislocations. Defective plastic deformations tend to occur in wider
crystalline sheets, where the emergent topological defects are anchored in space
serving as the seeds for subsequent fracture events. The fracture processes of
crystalline sheets of varying geometry are characterized by the sequences of
defining plastic events; the results are summarized in the phase diagram. The
effects of noise and initial lattice orientation on the plastic deformation of
the stretched crystalline sheet are also discussed. These results advance our
understanding of the atomic level on the plastic instabilities of 2D crystals
under large uniaxial stretching and may have potential practical implications in
the precise engineering of structural instabilities in packings of covalently
bonded particulate systems.

\begin{figure}[t] 
	\includegraphics[scale=0.23]{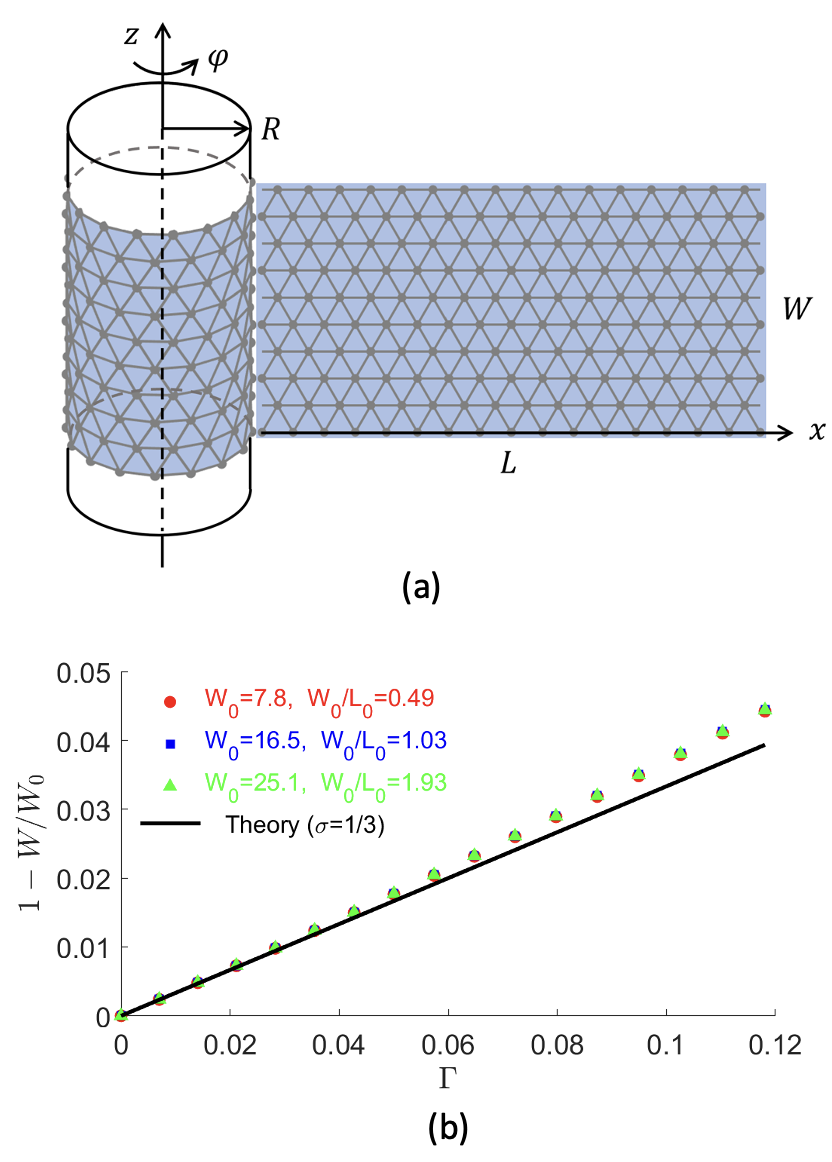}
  \caption{The schematic plot of the model system and the preliminary
  theoretical analysis. (a) The model system consists of a crystalline elastic
  sheet wrapping a cylinder under gradual expansion. The expanded crystalline
  sheet is also shown in the $x$-$z$ plane.  (b) Plots of $1-W/W_0$ against the expansion factor
  $\Gamma$ in the elastic deformation regime for the crystalline sheet of
  typical geometries. $W$ and $W_0$ are the width of the crystalline sheet
  before and after the deformation. The simulation data could be well
  fitted by the theoretical lines based on the linear elasticity theory up to
  $\Gamma=0.04$. }
	\label{model}
\end{figure}

\section{Model and Method}

The model system consists of the crystalline sheet wrapping a cylinder, as shown
in Fig.~\ref{model}(a). The crystalline sheet is fabricated by the L-J particles
in triangular lattice. The cylindrical substrate provides the geometric
constraint; there is no friction between the crystalline sheet and the cylinder.
The upper and lower edges of the crystalline sheet are free of external stress.
No external constraints other than the contact with the frictionless cylindrical
surface are imposed to the crystalline sheet. The particle interact by the
Lennard-Jones potential:
\begin{eqnarray}
	V(r_{ij})=4 \varepsilon_0
  \left[\left(\frac{\sigma_0}{r_{ij}}\right)^{12}-\left(\frac{\sigma_0}{r_{ij}}\right)^{6}\right],	
\end{eqnarray}
where $r_{ij}$ is the distance of particles $i$ and $j$ in three-dimensional
space. The L-J potential curve reaches the minimum value
$V_{min}=-\varepsilon_0$ at the balance distance of $r_0=2^{1/6}\sigma_0$.

In the initial state, the stress-free crystalline sheet of length $L_0$ and
width $W_0$ wraps on the cylinder of radius $R_0$. $L_0=2\pi R_0$.  The cylinder
is subject to controllable gradual expansion at a given rate.  Specifically, the
radius of the cylinder is enlarged by a factor of $1+p$ in each expansion. In
simulations, the parameter $p$ is sufficiently small ($p=0.7\%$) to fulfill the
quasi-static condition.  The particle configuration is subsequently relaxed to
the lowest energy state after each expansion, which is realized by the
standard steepest descent method in simulation~\cite{snyman2005practical}. The
step size $s=5\times 10^{-4}$.  In the relaxation of the stretched crystalline
sheet on the expanded cylinder of a given radius, the energy of the system is
reduced deep to the energy valley typically after $10^5$ updates of the particle
configuration.  After the $n$-th expansion, the radius of the cylinder becomes
$R_n = (1+p) R_{n-1}$. The total expansion factor after $n$
expansions is
\begin{eqnarray}
	\Gamma_n=\frac{R_n-R_0}{R_0}
\end{eqnarray}
The resulting particle configurations in mechanical equilibrium are analyzed
from both geometric and topological perspectives, including the variations of
the tilt angle and bond length, and the topological transformations of the
lattice via the Delaunay triangulation in the plastic deformations of the
stretched crystalline sheet~\cite{nelson2002defects}. In this work, the units of
length and energy are chosen to be the parameters $r_0$ and $\varepsilon_0$
associated with the L-J potential, respectively. No cut-off length of the L-J
potential is introduced in simulations.

\section{Results and discussion}

This section consists of four subsections. In Sec. III A, we perform
preliminary theoretical analysis on the homogeneous deformation of the
crystalline sheet under small stretching. In Sec. III B, we discuss the
defect-free intermittent plastic shear deformations upon large stretching.
Specifically, we analyze the lattice tilting phenomenon and establish its
connection to plastic shear deformations quantitatively based on a geometric
model. We also analyze the variation of bond length, the displacement field,
and the associated glide motion of dislocations in plastic shear deformations.
In Sec. III C, we discuss a distinct fracture mechanism based on the
proliferation of topological defects. The defective plastic deformations tend to
occur in wider crystalline sheets, where the emergent topological defects are
anchored in space serving as the seeds for subsequent fracture events. The
fracture processes of crystalline sheets of varying geometry are characterized
by the sequences of defining plastic events. The results are summarized in the
phase diagram. In Sec. III D, we discuss the effects of noise and initial
lattice orientation on the plastic deformation of the stretched crystalline
sheet.

\subsection{Homogeneous elastic deformation under small expansion}

The crystalline sheet consisting of L-J particles in triangular lattice, which
is confined on the cylindrical geometry, is initially stress free. With the
gradual expansion of the cylinder, the in-plane stress accumulates over the lattice.
Here, we first resort to the continuum elasticity theory to analyze the small
deformation of the crystalline lattice on the early stage of the expansion
process. By comparing with the numerical results, we also check the reliability
of the computational approach, which shall be used to explore the interested
regime of large deformation.

The triangular lattice is modeled as a two-dimensional continuous and isotropic
elastic sheet~\cite{Landau1986}. Due to its zero Gaussian curvature, the
cylindrical surface could be isometrically mapped to the plane~\cite{struik88a}.
This allows us to solve for the in-plane strain field in the Cartesian
coordinates $\{x, z\}$, as shown in Fig.~\ref{model}(a). $x=R\varphi$, where
$\varphi$ is the polar angle in the cylindrical coordinates. Here, for
convenience, the vertical axis in the Cartesian coordinates is also denoted as
$z$.

Now, we derive for the strain field over the stretched crystalline sheet in
mechanical equilibrium from the force balance equation $\partial_i \sigma_{ij} =
0$, where $i,j=x,z$, and by the following stress-strain relation for the
longitudinal deformation of the elastic membrane on the $x$-$z$
plane~\cite{Landau1986}
\be
    \sigma_{xx}&=&\frac{E}{1-\sigma^2}(u_{xx}+\sigma u_{zz}), \nonumber \\
    \sigma_{zz}&=&\frac{E}{1-\sigma^2}(u_{zz}+\sigma u_{xx}), \nonumber \\
    \sigma_{xz}&=&\frac{E}{1+\sigma}u_{xz},
	\ee
where $\sigma$ is the Poisson's ratio, and $E$ is the Young's modulus.
As a boundary condition, $\sigma_{zz}=0$ on the upper and lower boundaries of
the rectangular crystalline sheet. The sheet is horizontally stretched by a
factor of $\Gamma$. We thus obtain the expressions for the strain field: 
\be
u_{xx} &=& \Gamma  \nonumber \\             
u_{zz} &=& -\sigma u_{xx}=-\sigma \Gamma   \nonumber \\      
u_{xz} &=& u_{zx}=0. \label{equ:uzz}
\ee
According to Eqs.(\ref{equ:uzz}), the strain field established in the gently
stretched crystalline sheet in mechanical equilibrium is homogeneous.
Furthermore, a circumferential stretch by the factor of $\Gamma$ leads to a
transverse compression by the factor of $\sigma\Gamma$.  Note that $\sigma=1/3$
for the 2D isotropic elastic medium composed of L-J particles in triangular
lattice~\cite{PhysRevA.38.1005,berinskii2017plane,chen2022geometry}. From
Eqs.(\ref{equ:uzz}), we obtain the only non-zero component of the stress tensor: 
\be
\sigma_{xx}= E \Gamma,\label{sigma_xx}
\ee  
which is independent of the Poisson's ratio.

To check the reliability of the computational approach, we examine the variation
of the width of the crystalline sheet in the expansion of the cylinder in
simulations, and compare the numerical and theoretical results. We first present
the analytical expression for the width of the crystalline sheet at the
expansion factor $\Gamma$ according to the second equation in Eqs.(\ref{equ:uzz}):
\be
1- \f{W}{W_0} = \sigma \Gamma.   \label{equ:W}
\ee
In Fig.~\ref{model}(b), we present the plots of $1-W/W_0$ against the expansion
factor $\Gamma$. It turns out that the simulation data for the crystalline
sheets of varying width $W_0$ and aspect ratio $W_0/L_0$ collapse on the same
series of dots. It indicates that the relation of $1-W/W_0$ and $\Gamma$ is
independent of the geometry of the crystalline sheet. The agreement of the
simulation data and the theoretical result in Fig.~\ref{model}(b) shows the
reliability of the computational approach and the validity of the linear
elasticity theory in describing the elastic deformation of the L-J crystalline
sheet at least up to $\Gamma=0.04$.

\subsection{Defect-free intermittent plastic shear deformations}

\begin{figure*}[htbp] 
	\centering
	\includegraphics[scale=0.2]{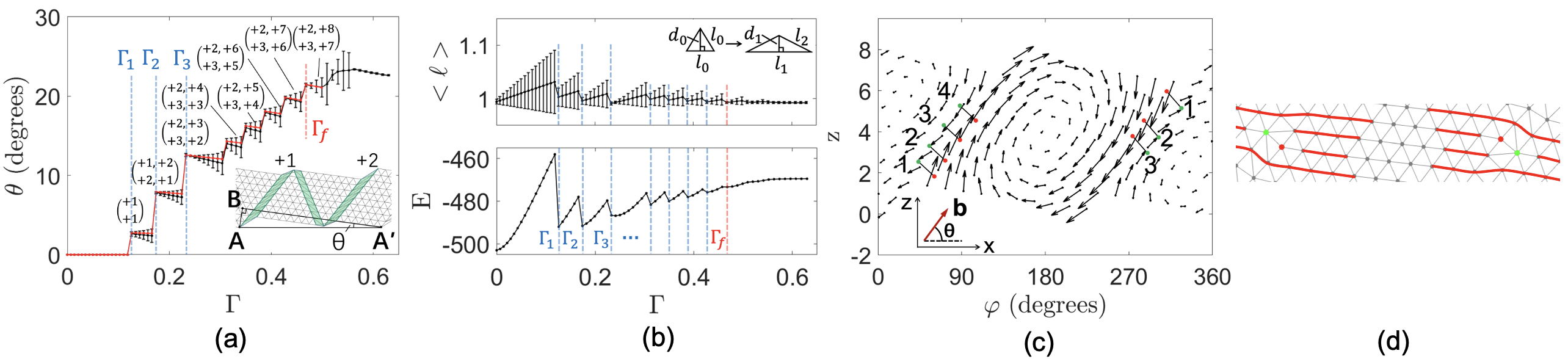}
  \caption{The adaptation of the crystalline sheet to the gradual expansion of
  the cylinder leads to the intermittent plastic shear deformations. The
  particle configurations in mechanical equilibrium in the stretching process
  are analyzed from the perspectives of the tilt angle (a), the bond length and
  energy (b), and the displacement vector field (c).  $\Gamma=0.22$ in (c). (d)
  Illustration of the pair of dislocations at the shear bands. The red and green
  dots represent five- and seven-fold disclinations.  $L_0=16.0$ and
  $W_0=7.8$.
	}
	\label{fig:tilt_angle}
\end{figure*}

\subsubsection{Shear-driven lattice tilting}


To explore the response of the crystalline sheet upon the gradual expansion of
the cylinder, we first analyze a typical example case presented in
Fig.~\ref{fig:tilt_angle}. An important observation is the tilting of the
strongly stretched crystalline lattice. The tilting phenomenon is
quantified in the plot of the tilt angle $\theta$ against the expansion factor
$\Gamma$ in Fig.~\ref{fig:tilt_angle}(a). $\theta$ is the angle between the
tilted lattice and the $x$ axis in the expanded planar crystalline sheet over the
$x$-$z$ plane. In the quasi-static expansion process,
the tilt angles of the bonds (with respect to $\vec{e}_{\varphi}$) in the
mechanically relaxed crystalline lattice at each value of $\Gamma$ are recorded.
The mean value and the standard deviation are plotted by the black curves with
error bars. The mean value of the tilt angle reflects the tilting of the entire
crystalline lattice.

A salient feature of the $\theta$-curve is that the variation of the tilt angle
$\theta$ exhibits the step-like behavior. With the expansion of the cylinder,
the value of $\theta$ is increased intermittently at a series of critical values
of $\Gamma$, which are denoted as $\Gamma_1$, $\Gamma_2$ ...  $\Gamma_f$ in
Fig.~\ref{fig:tilt_angle}(a). At $\Gamma=\Gamma_f$, the crystalline sheet is
completely fractured, meaning that the sheet wrapping the cylinder becomes
disconnected. As a practical criterion, the crystalline sheet wrapping the
cylinder is regarded as being completely fractured when the width of the
narrowest neck of two detaching lattice patches is less than about two lattice
spacings. On the early stage of the expansion process ($\Gamma < \Gamma_1$), the
crystalline lattice is subject to homogeneous deformation, and the tilt angle
remains zero until $\Gamma$ reaches the first critical value $\Gamma_1$;
$\Gamma_1 \approx 0.12$ for the case in Fig.~\ref{fig:tilt_angle}(a). Between
two consecutive critical values of $\Gamma$, it is noticed that the tilt angle
is slightly declined. It suggests that the lattice tends to tilt back to fit the
expanding cylinder. The tilt-back mode is energetically favored in comparison
with the pure stretching mode, which can be qualitatively understood by
considering a tilted straight 1D lattice with respect to the horizontal
reference line. By tilting back towards the reference line (even without
stretching the 1D lattice), the projected length of the 1D lattice along the
reference line is increased.

We examine the particle configurations in mechanical equilibrium in the
expansion process and find that the crystalline lattice experiences mechanical
instability in the form of irreversible plastic shear deformation at each
critical values of $\Gamma$. A typical plastically sheared particle
configuration is shown in the inset of Fig.~\ref{fig:tilt_angle}(a). The shear
bands resulting from the plastic deformation are highlighted in green.  These
observations suggest that the entire plastic deformation process of the crystalline sheet
could be characterized by the series of symbols $\Gamma_i$, each of which
represents a plastic event in the expansion of the crystalline sheet.

In the following, we shall show that the orientation of the shear band is
determined by the maximization of the shear stress as well as the orientation of
the crystalline lattice. Over the expanded crystalline sheet on the $x$-$z$
plane as shown in Fig.~\ref{model}(a), the uniaxial stretching along the
$x$ axis induces the non-zero component of the stress tensor $\sigma_{xx}$
in Eq.(\ref{sigma_xx}); $\sigma_{zz}=0$ due to the stress-free boundary
condition. To derive for the direction along which the shear stress reaches
maximum, we rotate the Cartesian coordinates $(x, z)$ counterclockwisely by angle $\alpha$ and work
in the rotated Cartesian coordinates $(x', z')$. For simplicity, 
the components of the vector $\vec{r}$ are denoted as $(x_1=x, x_2=z)$ and $(x_1'=x',
x_2'=z')$ in the original and rotated coordinates, respectively. These components
are related by the following relation:
\be
x_i' = A_{ij} x_j.
\ee
$A_{ij}$ is the component of the rotation matrix $\bf{A}$:
\be
\bf{A} = 
\lb \begin{array}{cc} \cos\alpha & \sin\alpha \\
           -\sin\alpha & \cos\alpha \end{array} \rb
\ee
The components of the stress tensor in the rotated Cartesian coordinates are:
\be
\sigma_{ij}'=A_{i\ell} A_{jm} \sigma_{\ell m},
\ee
in which the orthogonality condition of the rotation matrix is used;
$\bf{A}^{-1}=\bf{A}^{T}$.

We finally obtain the shear stress in the $(x', z')$ coordinates as:
$\sigma'_{12} = -\f{1}{2}E\Gamma \sin(2\alpha)$, whose magnitude reaches maximum
at $\alpha =\pi/4$ (and $\alpha = 3\pi/4$). Therefore, the crystalline sheet is
subject to maximum shear stress along the oblique line that makes the angle
$\pi/4$ with respect to the $x$ axis. In other words, the $\pi/4$ angle
represents the direction along which the applied uniaxial stretching is most
effectively converted into shear stress. 
Here, we shall point out that the orientation of maximum shear stress originates
from the transformation of the applied uniaxial stress tensor, and it is
independent of the microscopic interaction among the constituents composing the
isotropic elastic medium. However, simulations show that the microscopic
crystalline structure further restricts the orientation of the shear band.  In
the plastic shear deformation of the crystalline lattice at the tilt angle of
$\pi/3$, as shown in Fig.~\ref{fig:tilt_angle}(a), the shear band is along the
inclined principal axis of the crystalline lattice whose angle with respect to
the $x$ axis ($\pi/3$) is closest to the theoretically predicted value of
$\pi/4$ based on the continuum model. Fig.~\ref{fig:tilt_angle}(a) shows that
subsequent successive fracture of the crystalline lattice is along the formed
shear bands, leading to the intermittent increase of the tilt angle $\theta$.
Furthermore, considering that the plastic shear deformation involves the
enlarged separation of adjacent particles, the formation of the shear band
structure may be inhibited if the particle-particle attraction increases with
distance, which is opposite to the L-J potential. To test this point, we perform
simulations with harmonic potential in the cases of $L_0=(10, 20, 30)$ and
$W_0=(10, 20, 30)$; the crystalline sheet consists of a triangular lattice of
linear springs. It turns out that under the harmonic potential, no shear band
appears in the stretched crystalline lattice (at least up to $\Gamma=0.5$).


In the inset of Fig.~\ref{fig:tilt_angle}(a), we see that steps are formed
symmetrically on the upper and lower boundaries of the lattice as a consequence
of the plastic shear deformation. The emergent step structure changes the
morphology of the lattice boundary and breaks the originally $C_k$ symmetry of
the system, where $k$ is the number of particles on the crystalline line along
the circumferential direction.  The lattice tilting could be attributed to these
steps, which will be discussed later. Here, the steps-caused undulation of the
lattice boundary is distinct from the undulation phenomenon driven by the
Asaro-Tiller-Grinfeld (ATG)
instability~\cite{asaro1972interface,grinfel1986instability,srolovitz1989stability,kohler2015relaxation}.
In the latter case, the wave-like undulation developed on the stressed boundary
of the 2D or 3D elastic medium results from the instability for initial
perturbations of short wavelength, and mass transport is involved in the ATG
instability. In the pure mechanical system of the crystalline sheet wrapping the
cylinder, the stepped undulation of the boundary spontaneously occurs in the
absence of any initial perturbation and surface diffusion.

Now, we characterize the observed step structure on the boundaries of the
stretched crystalline sheet.  The state of the plastically sheared crystalline
lattice could be described by a series of numbers, each of which indicates the
height and direction of a step.  For example, the numbers above the particle
configuration in the inset of Fig.~\ref{fig:tilt_angle}(a) indicate the height
of the two steps on the upper boundary (in the unit of lattice spacing).  The
positive sign is to indicate an upward step (in the top view of the system in
the counterclockwise direction).  In Fig.~\ref{fig:tilt_angle}(a), the two
arrays of numbers in the bracket indicate the heights of the steps located on
the upper and lower boundaries, respectively. The dynamics of the steps is
constructed by tracking the variation of these numbers.  Specifically, on the
$\theta$-curve in Fig.~\ref{fig:tilt_angle}(a), we see that as the value of
$\Gamma$ exceeds $\Gamma_3$, the numbers in both columns increase by one,
indicating that the height of the two pairs of steps on both sides of the
lattice simultaneously increases by one lattice spacing. At the following larger
critical values of $\Gamma$, the numbers on the right column increase by one
or two each time, and those on the left column are invariant. It indicates that
a pair of steps rise by one or two lattice spacings each time, and the height
of the other pair of steps remains invariant in the expansion process.

Geometric analysis shows that the formation of the step structure on the
boundaries of the crystalline sheet caused by the plastic shear deformations is
the key to the observed lattice tilting. In the following, we shall show that
the total number of steps determines the tilt angle, regardless of the specific
distribution of the steps on the boundaries.

Our calculations are based on the typical plastically sheared lattice in the
inset of Fig.~\ref{fig:tilt_angle}(a). To measure the tilt angle, we first
connect the two identical points $A$ and $A'$ by a horizontal line on the plane
of the unfolded cylindrical surface. Along the lattice line, we draw the line
$A'B$ and make the right triangle $\bigtriangleup ABA'$. All of the steps on the
lower boundary of the lattice are enclosed in the triangle $\bigtriangleup
ABA'$. The tilt angle $\theta$ is the angle between the lines $A'A$ and $A'B$.
Based on this geometric model, geometric arguments show that  
\begin{eqnarray}
	\sin\theta = \frac{\sqrt{3}}{4\pi}\frac{ \ell(\Gamma) }{R(\Gamma)}	N_{s},
	\label{Eq:theta}
\end{eqnarray} 
where $N_{s}$ is the total number of steps enclosed in the triangle
$\bigtriangleup ABA'$, $\ell$ is the bond length, and $R$ is the radius of the
cylinder. Simulations show that the relative variations of the bond length in
the stretched lattices are small in comparison with the mean bond length even in
the strongly stretched regime. As such, it is assumed that the bond length takes
a uniform value $\ell(\Gamma)$ in Eq.(\ref{Eq:theta}). Equation(\ref{Eq:theta})
shows that $\sin\theta$ is proportional to the total number of steps $N_{s}$,
and its dependence on the expansion factor $\Gamma$ is through the quantities
$\ell(\Gamma)$ and $R(\Gamma)$.

According to Eq.(\ref{Eq:theta}), the tilt angle $\theta$ versus the expansion
factor $\Gamma$ is plotted by the red curves in Fig.~\ref{fig:tilt_angle}(a). To
obtain the $\theta$-curve, at each value of $\Gamma$, we calculate the mean bond
length of the stretched lattice in mechanical equilibrium for the value of the
quantity $\ell(\Gamma)$, and count the total number of steps $N_{s}$.
Figure~\ref{fig:tilt_angle}(a) shows that the numerical results (the black curve
with error bars) are well fitted by the red curve based on the geometric model
in Eq.(\ref{Eq:theta}). The small deviation of the black and red curves between
consecutive critical values of $\Gamma$ can be attributed to the overestimated
value of $\ell(\Gamma)$ in Eq.(\ref{Eq:theta}). Specifically, the mean bond
length that is used in the plot of the theoretical curve is slightly larger than
the actual length of the bonds along the direction of the line $AB$ in the inset
of Fig.~\ref{fig:tilt_angle}(a) due to the Poisson effect.
Overall, the agreement of the numerical and theoretical results shows the
validity of the geometric model for quantitatively understanding the tilting
phenomenon caused by the plastic shear deformations.


In preceding paragraphs, we show that the crystalline sheet adapts to the
expanding cylindrical substrate via the stretching and the tilting modes for the
case in Fig.~\ref{fig:tilt_angle}. Specifically, prior to the occurrence of the
first plastic shear deformation, the response of the lattice is to stretch the
bond length in the circumferential direction and simultaneously squeeze the
lattice in the perpendicular direction; the tilt angle remains invariant. The
adjustment of the bond length conforms to the mechanical laws of continuum
elasticity theory. Once the expansion factor $\Gamma$ reaches a series of critical
values, the adaptation of the crystalline sheet to the expanding cylinder is
through the tilting of the entire lattice, which is realized by the plastic
shear deformations. The connection of the tilting phenomenon and the plastic
shear deformations of the strongly stretched crystalline lattice is established
based on the geometric model.

\subsubsection{Geometric analysis of bond length and displacement field}


We proceed to analyze the variation of the bond length in the plastic shear
deformation of the case in Fig.~\ref{fig:tilt_angle}(a). In the upper panel in
Fig.~\ref{fig:tilt_angle}(b), the mean bond length $\la \ell\ra$ is plotted
against the expansion factor $\Gamma$. The standard deviation of the bond length
distribution is indicated by the error bars. Note that the bonds on the
boundary are excluded in the statistical analysis of the bond length.

From the upper panel in Fig.~\ref{fig:tilt_angle}(b), we see that the $\la
\ell\ra$-$\Gamma$ curve exhibits the zig-zag behavior. In the expansion process
up to $\Gamma \approx 0.5$, the $\la \ell\ra$-$\Gamma$ curve conforms to the
following pattern: the value of the mean bond length increases linearly with the
expansion factor $\Gamma$ and then rapidly falls down. In this process, the
magnitude of the variation of $\la \ell\ra$ is reduced with the increase of
$\Gamma$. Scrutiny of the particle configurations in mechanical equilibrium
indicates that each turning point of the $\la \ell\ra$-$\Gamma$ curve exactly
corresponds to the plastic shear deformations of the lattice. The series of the
critical values of $\Gamma$ in Fig.~\ref{fig:tilt_angle}(b) are identical to
those in Fig.~\ref{fig:tilt_angle}(a), where the plastic shearing events are
analyzed in terms of the tilt angle. Examination of the variation of the energy
reveals the abrupt decline of the energy at these critical values
of $\Gamma$, as shown in the lower panel of Fig.~\ref{fig:tilt_angle}(b).

From the $\la \ell\ra$-$\Gamma$ curve in Fig.~\ref{fig:tilt_angle}(b), we also
see that the standard deviation of the bond length distribution (indicated by
the error bars) increases prior to each plastic shear deformation. Theoretical
and numerical analysis shows that this phenomenon can be attributed to the
Poisson effect.  Qualitatively, as the crystalline sheet is stretched along one
direction ($u_{xx}>0$), it shrinks in the perpendicular direction
($u_{zz}<0$), enlarging the dispersion in the distribution of the bond length.
This effect is illustrated in the inset of the upper panel in
Fig.~\ref{fig:tilt_angle}(b), where the left and right triangles represent an
elementary cell in the triangular lattice before and after the deformation.

Here, we present a quantitative geometric analysis based on the inset in
Fig.~\ref{fig:tilt_angle}(b) to explain the observed variation of the error bars
on the $\la \ell\ra$-$\Gamma$ curve. In the deformed triangle, $\ell_2=\sqrt{(\ell_1/2)^2+d_1^2}$, where $\ell_1=(1+\Gamma)\ell_0$.
By the definition of Poisson's ratio $\sigma=-u_{zz}/u_{xx}$, we have
$d_1=(1-\sigma \Gamma)d_0$, where $\sigma$ is the Poisson’s ratio. We finally
obtain an upper bound estimation for the dispersion of the bond length
distribution: 
\be
\f{\delta \ell}{\ell_0} &=& \f{\ell_1-\ell_2}{\ell_0}  \nonumber \\
&=& 1+\Gamma - \sqrt{1+ \frac{1-3\sigma}{2}\Gamma+ \frac{1+3\sigma^2}{4}\Gamma^2} 
\nonumber   \\
&=& \frac{3}{4}(1+\sigma)\Gamma  -\frac{3}{32}(1+\sigma)^2\Gamma^2 + O(\Gamma^4). 
\label{delta_ell}
\ee
For $\sigma=1/3$, $\delta \ell/\ell_0=\Gamma - \Gamma^2/6 +O(\Gamma^4)$, where
the coefficient in the quadratic term is much less than that in the linear term.
As such, the dispersion of the bond length distribution increases approximately
linearly with $\Gamma$. This is in agreement with the numerical results in
Fig.~\ref{fig:tilt_angle}(b), where the relative reduction of the slope of
$\delta \ell(\Gamma)$ is less than $5\%$ in the range of $\Gamma \in [0,
\Gamma_1]$.  The preceding argument based on the deformation of the elementary
triangle could be extended to the case of a tilted lattice in mechanical
equilibrium for $\Gamma > \Gamma_1$ in Fig.~\ref{fig:tilt_angle}(b). The
conclusion remains valid that the dispersion of the bond length distribution is
enlarged with the expansion of the cylinder.


Now, we analyze the plastic shear deformation from the perspective of the
displacement field, as shown in Fig.~\ref{fig:tilt_angle}(c). The displacement
field is constructed based on the two particle configurations in mechanical
equilibrium on the cylinders of radii $R$ and $R+\delta R$ in an expansion.
Specifically, we project each particle in the latter particle configuration
radially (along the direction of $-\hat{e}_r$) onto the cylinder of radius $R$.
The difference of the projected particle configuration and the original one on
the cylinder of radius $R$ yields the displacement field.
Fig.~\ref{fig:tilt_angle}(c) shows that the displacement field in a shear event
is featured with the vortex structure. The coherent displacements of the
particles in the vortex region are driven by the anti-parallel displacement
vectors along the two adjacent shear bands. The shear-driven emergence of
vortices, which is widely seen in fluids, has been reported in a compressed 2D
lattice system~\cite{sun2024plastic}. Here, we show that a stretched lattice
also supports the vortex structure under shear deformation, indicating the
generality of the scenario of shear-driven vortex structure in solid mechanical
systems.

\subsubsection{Glide of dislocations in intermediate states}


In the mechanical relaxation process for the case presented in
Fig.~\ref{fig:tilt_angle}, dislocations are observed within the shear band in
the intermediate states; the intermediate states refer to the particle
configurations in the mechanical relaxation process of the system at a given
value of the expansion factor $\Gamma$. The remaining region of the crystalline
sheet is free of topological defects. The results are recorded in
Fig.~\ref{fig:tilt_angle}(c).  A dislocation consists of a pair of five- and
seven-fold disclinations, which are indicated by the red and green dots in
Fig.~\ref{fig:tilt_angle}(c). A $p$-fold disclination refers to a point whose
coordination number is equal to $p$, and it represents a fundamental topological
defect in the triangular lattice~\cite{nelson2002defects}.

Figure~\ref{fig:tilt_angle}(c) shows that on the early stage of the relaxation
process, a pair of dislocations labeled 1 simultaneously emerge on the adjacent
shear bands.  The subsequent migration of the dislocations is indicated by the
labels 2, 3 and 4. The values of the simulation steps at the labels from 1 to
4 are $n_s=27,000$, $29,500$, $30,500$, and $31,000$, respectively. Here, the
numerical experiment at a fine simulation step allows us to capture the
anti-parallel glide motion of the pair of the dislocations along the shear
bands, which facilitates the shear deformation.

To understand the anti-parallel glide motion of the dislocations, we examine the
force on the pair of the dislocations.  The presence of the dislocation causes
the deformation of an originally perfect crystal. In the small panel in
Fig.~\ref{fig:tilt_angle}(c), we show the deformed crystal lattice near a
dislocation (the pair of green and red dots) on the shear bands. By the lines
along the crystal lattice, we show that the effect of the dislocation is to
insert an extra array of particles. In the continuum elasticity description, the
property of the dislocation is characterized by the contour integral of the
resulting displacement field enclosing the dislocation
concerned~\cite{Landau1986}. It leads to the Burgers vector $\vec{b}$ whose
magnitude is equal to the lattice spacing. The Burgers vector is a topological
quantity in the sense that it is independent of the geometry of the contour.

The Burgers vector specifies the direction of the glide motion of the
dislocation~\cite{hirth1982theory}. In the inset of
Fig.~\ref{fig:tilt_angle}(c), the Burgers vector $\vec{b}$ associated with the
dislocation on the left shear band is plotted on the $x$-$z$ plane.  $\vec{b}$
is perpendicular to the connecting line of the five- and seven-fold
disclinations and it makes an angle $\phi$ with respect to the $x$ axis. The
signs of the Burgers vectors associated with the pair of dislocations on the
adjacent shear bands are opposite.

\begin{figure}[t] 
	\includegraphics[scale=0.21]{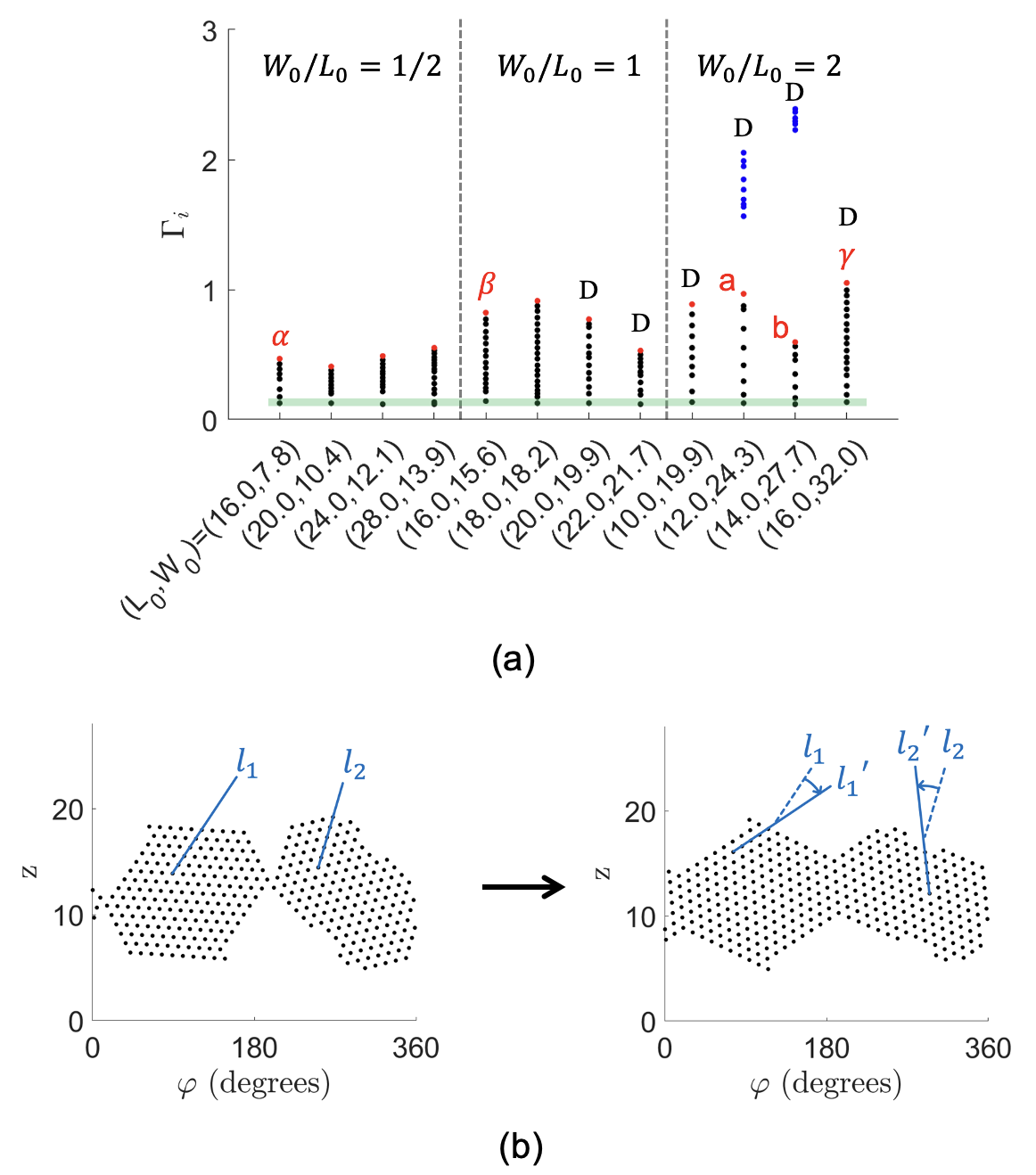}
  \caption{Characterization of the intermittent plastic deformation process.
  (a) Plot of the critical values of $\Gamma$ for the crystalline sheets of
  varying size and aspect ratio. The critical value of $\Gamma$ is recorded for
  each plastic event in the expansion process. The complete fracture
  (disconnection) of the crystalline sheet wrapping the cylinder occurs at the
  terminal dots (in red). In the cases marked by letter D, topological defects
  remain in the mechanically relaxed lowest energy states. (b) The
  rotate-and-reconnect phenomenon after the complete fracture of the crystalline
  sheet. The value of $\Gamma$ is increased from $\Gamma=1.547$ to
  $\Gamma=1.564$ in the two particle configurations. The post-fracture critical
  values of $\Gamma$ are indicated by the blue dots in (a). }
	\label{fig:Gamma_n}
\end{figure}

\begin{figure*}[th] 
	\centering
	\includegraphics[scale=0.43]{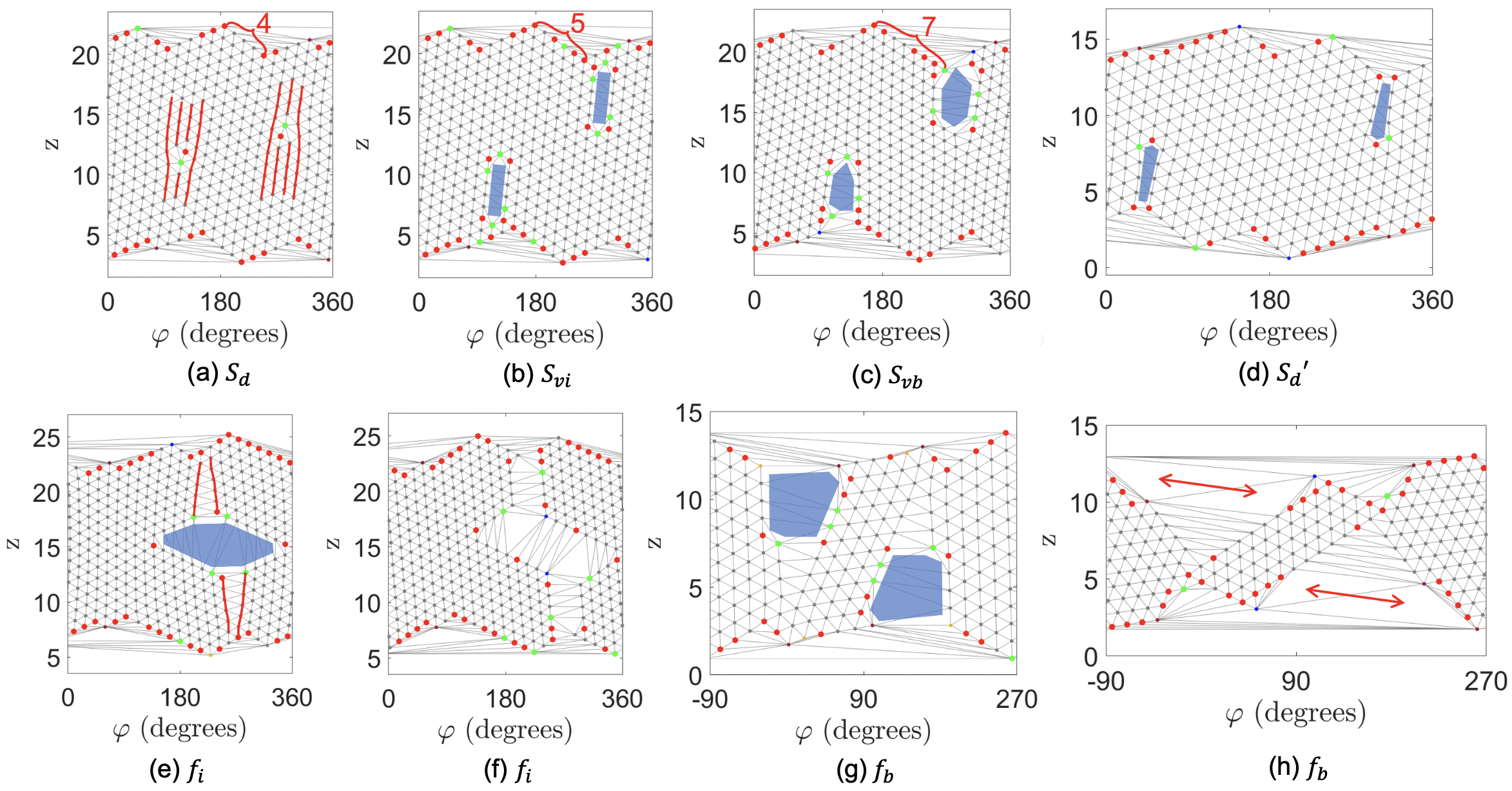}
  \caption{Characteristic defect structures define the stable
  states in the fracture process of the crystalline sheet. (a)-(c) The fracture
  process in the example case of $L_0=13.0$ and $W_0=25.1$ follows the sequence of
  $S_{d} \rightarrow S_{vi} \rightarrow S_{vb}$, where the symbols refer to the defect states defined by
  the emergence of isolated dislocations (a), elongated vacancies in the
  interior of the lattice (b), and fractures at the boundary (c). (d) The defect
  state $S_{d'}$ is characterized by the presence of isolated dislocations and
  the associated strips of square plaquettes (highlighted in blue) in the
  background of the triangular lattice. (e)-(f) and (g)-(h) show the fracture
  modes $f_i$ and $f_b$ under which the disconnection of the sheet is initiated
  from the interior and the boundary of the sheet, respectively.
	}
	\label{fig:defective}
\end{figure*}

The dislocation that is characterized by the Burgers vector $\vec{b}$ is subject
to the Peach-Koehler force in the external stress field
$\sigma_{ij}^{(e)}$~\cite{peach1950forces}:
\be
f_i = -\varepsilon_{ij}\sigma_{jk}^{(e)}b_k, \label{peach}
\ee 
where $\varepsilon_{ij}$ is the anti-symmetric tensor. The Burgers vector on the
$x$-$z$ plane is subject to the stress field in Eq.(\ref{sigma_xx}). From
Eq.(\ref{peach}), we obtain the $z$ component of the Peach-Koehler force: $f_z =
E\Gamma b_x$. The $x$ component of the force is zero. Therefore, the component
of the Peach-Koehler force along the direction of the Burgers vector is 
\be
f_{\vec{b}} &=& f_z \cos\lb \f{\pi}{2} - \phi \rb \nonumber \\
&=& \f{1}{2}E\Gamma b \sin(2\phi),  \label{fb}
\ee
which reaches maximum at $\phi=\pi/4$ or $\phi=5\pi/4$.
Eq.(\ref{fb}) shows that $f_{\vec{b}}$ is invariant in the transformation of
$\phi \rightarrow \phi +\pi$. It indicates that the Peach-Koehler forces
along the Burgers vectors of opposite signs are anti-parallel. This explains the
observed anti-parallel glide motion of the dislocations along the adjacent shear
bands in Fig.~\ref{fig:tilt_angle}(c). 

Here, it is of interest to note that dislocations in cylindrical crystals
exhibit rich phenomena despite of the zero Gaussian curvature of the cylindrical
surface~\cite{amir2013theory}. Besides the glide motion parallel to the Burgers
vector, dislocations also climb perpendicular to the Burgers vector; the growth
of the cell walls of rod-shaped bacteria may be regarded as mediated by
dislocation climb~\cite{hirth1982theory}. Furthermore, in tubular crystal, which
is modeled by the deformable network of harmonic springs in cylindrical topology,
dislocation glide and its connection to the reconfiguration of the lattice in
terms of parastichy transition have been investigated by the combination of
analytical continuum elasticity theory and numerical
simulations~\cite{beller2016plastic}. In Ref.~\cite{beller2016plastic}, the
stress is applied along the axis of the tube, and the dislocation glide is realized by
bond flip in the triangular lattices of spring bonds. The distinct physical
mechanisms underlying the glide of dislocations in the lattices of spring bonds
and L-J particles share a common topological consequence. Specifically, dislocation
glide leads to the changes in the parastichy numbers in the former case and
the stepwise transformations in the latter case; the reason is that the effect
of a dislocation in a triangular lattice is to insert an array of
particles~\cite{Landau1986}.

\subsubsection{Characterization of the intermittency of plastic deformations}


We systematically investigate the plastic deformations in the expansion of
crystalline sheets of varying geometry. The value of $\Gamma$ is recorded for
each plastic event in the expansion process, including the formation of steps on
the boundary of the sheet and the relocation of particles in the interior of the
lattice. The results for crystalline sheets of typical aspect ratios are grouped
and presented in Fig.~\ref{fig:Gamma_n}(a). In Fig.~\ref{fig:Gamma_n}(a), the
series of the critical values of $\Gamma$ as represented by the columns of dots
indicate the intermittent nature of the plastic deformations of the stretched
crystalline sheet under the gradual expansion of the cylinder. Note that in the
cases marked by letter D in Fig.~\ref{fig:Gamma_n}(a), topological defects
remain in the mechanically relaxed lowest energy states. The defect-based
fracture processes will be discussed in Sec. III C.

From Fig.~\ref{fig:Gamma_n}(a), we observe that the first critical values
$\Gamma_1$, at which the originally crystalline lattice is subject to initial
plastic deformation, are located within the thin bar in green and they are
insensitive to the size and aspect ratio of the crystalline sheet.  Statistical
analysis of the data presented in Fig.~\ref{fig:Gamma_n}(a) shows that for the
crystalline sheets of varying size and aspect ratio, $\Gamma_1$ takes a
relatively uniform value: $\Gamma_1 = 0.126 \pm 0.008$. In contrast, the
values of $\Gamma_f$ as indicated by the red dots in Fig.~\ref{fig:Gamma_n}(a),
at which the crystalline sheet is completely fractured, exhibit appreciable
discrepancy not only for the crystalline sheets of different aspect ratios but
also for those of different sizes and identical aspect ratio.

Here, we compare the crystalline sheets $\alpha$, $\beta$ and $\gamma$ of
identical length $L_0$ and varied width $W_0$ in Fig.~\ref{fig:Gamma_n}(a) and
find that the value of $\Gamma_f$ is larger for a wider sheet. It suggests that
increasing the width of the crystalline sheet enhances its ability to resist
fracture upon the expansion. It is noticed that in general the dots in the
columns are unevenly distributed, implying the sensitivity of the critical
values of $\Gamma$ to the stretching deformation for the highly stretched
crystalline sheet. The enhanced sensitivity of the crystalline sheet under
stronger stretching is confirmed by checking the example case of $(L_0, W_0)=
(16.0, 15.7)$ at varying step size. Specifically, as the value of the step size
is varied in the range from $s=5\times 10^{-4}$ to $s=10^{-3}$, the critical
value of $\Gamma_1$ is varied in the narrow range of $\Gamma_1 = 0.129 \pm
0.008$. In contrast, the critical value of $\Gamma_f$ is subject to a larger
fluctuation of $\Gamma_f = 0.749 \pm 0.228$, which could be attributed to the
enhanced sensitivity of the highly stretched crystalline sheet prior to complete
fracture. Note that the intermittent nature of the plastic shear deformations is
unchanged with the variation of the step size.


The numerical approach also captures post-fracture events. Specifically, in the
case of $(L_0, W_0)=(12.0, 24.3)$ [labeled by $a$ in Fig.~\ref{fig:Gamma_n}(a)],
the two crystal patches connected by a neck whose width is as thin as one
lattice spacing (which is regarded as being disconnected by our practical
criterion) are observed to rotate and reconnect neatly, as shown in
Fig.~\ref{fig:Gamma_n}(b); the value of $\Gamma$ is increased by $1.1\%$ (from
$\Gamma=1.547$ to $\Gamma=1.564$). In this process, the rotational motion of the
crystal patches along the opposite directions, which can be excited without
costing much energy, is the key to triggering the energetically favored
reconnection event.  Similar process is observed in the case of $(L_0,
W_0)=(14.0, 27.7)$ [labeled by $b$ in Fig.~\ref{fig:Gamma_n}(a)]. It seems that
the crystalline sheet exhibits a strong tendency to reconnect even if the
disconnection has already occurred, all out of the attractive nature of the L-J
potential in the stretched regime.  Further stretching the stable particle
configuration in the reconnected state in Fig.~\ref{fig:Gamma_n}(b) leads to a
series of plastic deformations; the corresponding critical values of $\Gamma$
are recorded in Fig.~\ref{fig:Gamma_n}(a) by the blue dots.

\subsection{Defect-based fracture mechanism}

\begin{table*}[]
	\begin{tabularx}{\textwidth}{|c|X|X|X|X|X|}
		\hline
		\multicolumn{1}{|c|}{\diagbox{$W_0$}{$L_0$}} & 10.0 & 15.0 & 20.0 & 25.0 & 30.0 \\ \hline
    10.4 & $S_0\to S_{vi}\to S_0\to f_s$ & $ \blue{S_0\to f_s}$ & $ \blue{S_0\to f_s}$ & $ \blue{S_0\to f_s}$ & $ \blue{S_0\to f_s}$ \\ \hline
		14.7 & $S_0\to S_d\to S_d’\newline \to S_{vi} \to S_{vb} \to f_b$ & $ \blue{S_0\to f_s}$ & $ \blue{S_0\to f_s}$ & $ \blue{S_0\to f_s}$ & $ \blue{S_0\to f_s}$ \\ \hline
		19.9 & $S_0\to S_{vi}\to f_s\&f_i$ & $S_0\to S_d\to S_0\to S_d\newline \to S_0\to S_{vb}\to f_s\&f_b$ & $S_0\to S_{vi}\to S_0\newline \to S_{vi}\to S_{vi}\&S_{vb}\newline \to f_s\&f_i\&f_b$ & $ \blue{S_0\to f_s}$ & $ \blue{S_0\to f_s}$ \\ \hline
		25.1 & $S_0\to S_{vb}\to f_s\&f_b$ & $S_0\to S_d\to S_0\to S_d\newline \to S_0\to S_d\to S_{vi}\newline \to S_{vb}\to f_s\&f_b$ & $S_0\to S_d\to S_0\to S_d’\newline \to S_{vb}\to f_s\&f_b$ & $ \blue{S_0\to f_s}$ & $ \blue{S_0\to f_s}$ \\ \hline
		30.3 & $S_0\to S_{vi}\to f_i\&f_s$ & $S_0\to S_d\to S_0\to S_d\newline \to S_d’ \to S_{vi}\to S_{vb}\newline \to f_b\&f_s$ & $S_0\to S_{vi}\to f_s\&f_i$ & $S_0\to S_{vi}\to S_0 \newline \to S_{vb}\to f_s\&f_b$ & $S_0\to S_q\to S_q\&S_{vb}\newline \to S_q\&S_{vb}\&S_{vi}\newline \to S_{vb}\&S_{vi}\newline \to S_{vb}\&S_q\to S_{vb}\newline \to S_0\to S_q\to S_{vi}\newline \to S_0\to S_d\to S_0\newline \to S_{vb}\to f_s\&f_b$ \\ \hline
	\end{tabularx}
  \caption{The phase diagram for the defect-free and defect-based fracture
  mechanisms, as distinguished by the blue and black font colors, respectively.
  The sequences show the stable states in the fracture process.  The symbols
  $S_{*}$ refer to the defect states defined by the emergence of isolated
  dislocations ($S_{d}$ and $S_{d'}$), quadrupoles ($S_q$) and elongated
  vacancies in the interior ($S_{vi}$) and on the boundary ($S_{vb}$) of the
  lattice. The defect-free state is indicated by $S_0$.  The disconnection
  behaviors of the crystalline sheet are characterized by the three kinds of
  modes: $f_i$ (via the interior fractures), $f_b$ (via the fractures on the
  boundary) and $f_s$ (via defect-free plastic shear
  deformation).
  }
  \label{phase_diagram}
\end{table*}

Systematic investigations of crystalline sheets of varying geometry shows that
topological defects may remain in the mechanically relaxed lowest energy states.
We shall show that these topological defects provide a fracture mechanism that
is distinct from the defect-free fracture process as discussed in the previous
subsection. In the latter process, dislocations vanish via the continuous slide
along the shear band.


We first examine the fracture process in a specific example case of $L_0=13.0$ and
$W_0=25.1$. The typical lattice configurations in mechanical equilibrium are
presented in Figs.~\ref{fig:defective}(a)-\ref{fig:defective}(c). The red and
green dots represent five- and seven-fold disclinations. With the expansion of
the cylinder, we first observe the emergence of isolated dislocations ($5-7$
disclination pairs) in the interior of the lattice and the simultaneous
formation of step structures on the boundaries of the lattice. In
Fig.~\ref{fig:defective}(a), by the lines along the crystal lattice, we show
that the opposite signs of the Burgers vectors associated with the two
dislocations correspond to the insertion of particle arrays along opposite
directions.

Increasing the value of $\Gamma$ from $\Gamma=0.44$
[Fig.~\ref{fig:defective}(a)] to $\Gamma=0.48$ [Fig.~\ref{fig:defective}(b)]
leads to the instability of the isolated dislocations, resulting in the
stretch-driven formation of elongated vacancies (interior fractures) as
highlighted in blue in Fig.~\ref{fig:defective}(b). This scenario is
fundamentally different from the glide of the dislocations in the crystalline
sheet of smaller width in Fig.~\ref{fig:tilt_angle}. Here, the dislocations
are anchored in space and they serve as the seeds for the subsequent vacancy
structures.  Under the gradual expansion of the cylinder, these interior
fractures are further torn apart and extended to the boundaries of the lattice
as shown in Fig.~\ref{fig:defective}(c). The characteristic defect states in
Figs.~\ref{fig:defective}(a)-\ref{fig:defective}(c) are denoted as $S_{d}$
(isolated dislocations), $S_{vi}$ (elongated vacancies in the interior of the
lattice) and $S_{vb}$ (elongated vacancies at the boundary), respectively. In
the expansion of the cylinder, we also notice the growth of the step structure
in its height; the step heights are indicated by the numbers.

The extension of the elongated vacancies in the defects states $S_{vi}$ and
$S_{vb}$ under the uniaxial stretching of the crystalline sheet could be
understood in the framework of the Griffith theory for crack
stability~\cite{griffith1921vi}.  According to the Griffith theory, the
propagation of an existing crack in a crystal occurs when the decrease in the
elastic strain energy exceeds the increase in surface energy created by the new
crack surface. The scenario of the crack propagation based on the continuum
Griffith theory has been developed in a series of atomic-scale simulations by
incorporating the discrete lattice structure of the
crystal~\cite{ippolito2006role}, the role of
dislocations~\cite{cheung1990brittle}, the strain-dependence of the Young's
modulus and surface energy as well as the effect of lattice
trapping~\cite{mattoni2005atomic}.  In our system, the observed extension of the
elongated vacancies at zero temperature is also driven by the reduction of
energy according to the steepest descent algorithm. It is of interest to carry
out detailed analysis of different energy contributions in the crack process; it
is beyond the scope of current investigation.

The fracture process of the crystalline sheet shown in
Figs.~\ref{fig:defective}(a)-\ref{fig:defective}(c) could be represented by the
sequence of $S_{d}\rightarrow S_{vi} \rightarrow S_{vb}$. This route of fracture
is distinct from defect-free fracture process in Fig.~\ref{fig:tilt_angle},
where topological defect emerge only in the intermediate states. Here, the
emergence of defects implies that the fracture mechanism based on pure shear
deformation like in the case of Fig.~\ref{fig:tilt_angle} becomes insufficient
for the crystalline sheet to withstand the expansion of the cylinder. A new
fracture mechanism based on the proliferation of defects is activated in a
highly stretched crystalline sheet to adapt to the expanding cylinder. Here, we
also report the observation of the strips of square plaquettes [highlighted in
blue in Fig.\ref{fig:defective}(d)] embedded in the triangular lattice in the
partial shear of the lattice; the particle configuration is in mechanical
equilibrium.

Upon the gradual expansion, the highly stretched crystalline sheet is ultimately
completely fractured. We examine the entire fracture processes of crystalline
sheets of varying geometry and find that the disconnection of the sheet follows
three kinds of modes. In the first mode, the crystalline sheet is completely
fractured via the continuous slide of the lattice along the shear band as shown
in Fig.~\ref{fig:tilt_angle}. This mode is denoted as $f_s$. In the second
and third modes, the disconnection of the sheet is initiated from the interior
and the boundary of the lattice, respectively. They are denoted as $f_i$ and
$f_b$, respectively.

Specifically, the $f_i$ mode is illustrated in Figs.\ref{fig:defective}(e) and
\ref{fig:defective}(f).  Under the longitudinal stretching, the eye-shaped hole
(highlighted in blue) is fractured approximately at the locations of
$\phi=\pi/2$ and $\phi=3\pi/2$ (marked by the red lines), where $\phi$ is
the polar angle.  These preferred fracture sites could be understood by the
model of the circular hole under the uniform horizontal tension $S$. According
to the continuum elasticity theory, the azimuthal component of the stress
tensor $\sigma_{\phi\phi}$ around the circular hole in an isotropic elastic
medium reaches maximum $(\sigma_{\phi\phi})_{\textrm{max} }=3S$ at
$\phi=\pi/2$ or $\phi=3\pi/2$~\cite{timoshenko1951theory,Landau1986}. In
Fig.\ref{fig:defective}(e), the vertical fractures are driven by the stress
$\sigma_{\phi\phi}$ and they tend to occur on sites where
$\sigma_{\phi\phi}$ reaches maximum. In other words, the fractures initiated
from the interior of the lattice could be attributed to the stress-focusing
effect around the preexistent vacancies.

In the $f_b$ mode as shown in Figs.\ref{fig:defective}(g) and
\ref{fig:defective}(h), the disconnection of the crystalline sheet is caused by
the extension of the pre-existent fractures on the boundary (highlighted in
blue).  Here, we shall point out that in some cases, the disconnection process
involves the combination of the three kinds of modes ($f_s$, $f_i$, and $f_b$).
For example, we observe the realization of the complete fracture via the
simultaneous slide along the shear band and the extension of the fracture on the
boundary. This disconnection mode is denoted as $f_s\& f_b$.


In Table~\ref{phase_diagram}, we list all of the stable states in the fracture
process of the stretched crystalline sheets of typical geometries upon the
gradual expansion of the cylinder. $S_0$ refers to the defect-free state. The
other symbols $S_{*}$ refer to the defect states defined by the emergence of
isolated dislocations ($S_{d}$ and $S_{d'}$), quadrupoles ($S_q$) and elongated
vacancies in the interior ($S_{vi}$) and on the boundary ($S_{vb}$) of the
lattice. The disconnection behaviors of the crystalline sheet are summarized by
the three kinds of modes: $f_i$, $f_b$ and $f_s$.

From the columns of the entries in Table~\ref{phase_diagram}, we see that
increasing the width of the crystalline sheet leads to the transition in the
fracture mechanism from the defect-free to the defect-based plastic
deformations, which are distinguished by the blue and black colors in the Table.
The transition in the fracture mechanism with the increase of $W_0$ suggests
that the coordinated movement of the particle array along the shear band in the
defect-free fracture process is ultimately disrupted along a sufficiently long
shear band. From the perspective of topological defects, the excited
dislocations induce the formation of elongated vacancies prior to gliding to the
boundary along a long shear band. Note that in Fig.~\ref{fig:Gamma_n} all of the
cases in the group of $W_0/L_0=2$ and some cases in the group of $W_0/L_0=1$
(indicated by letter D) belong to the category of defect-based fracture and the
remaining cases (with $W_0/L_0$ being equal to one or less) follow the
defect-free fracture mode, which is consistent with the results presented in
Table~\ref{phase_diagram}. Here, we shall point out that even in the
defect-free fracture mechanism as shown in Fig.~\ref{fig:tilt_angle},
dislocations arise in the intermediate states to facilitate the plastic shear
deformation via the glide motion; these defects ultimately vanish in the
mechanically relaxed states in crystalline sheets of short width. From
Table~\ref{phase_diagram}, we also notice that the crystalline sheet tends to be
disconnected from the interior (in the $f_i$ mode) with the appearance of
interior vacancies (in the defect state $S_{vi}$); otherwise, the disconnection
occurs from the boundary of the sheet (in the $f_b$ mode).

\begin{figure}[t] 
	\centering
	\includegraphics[scale=0.25]{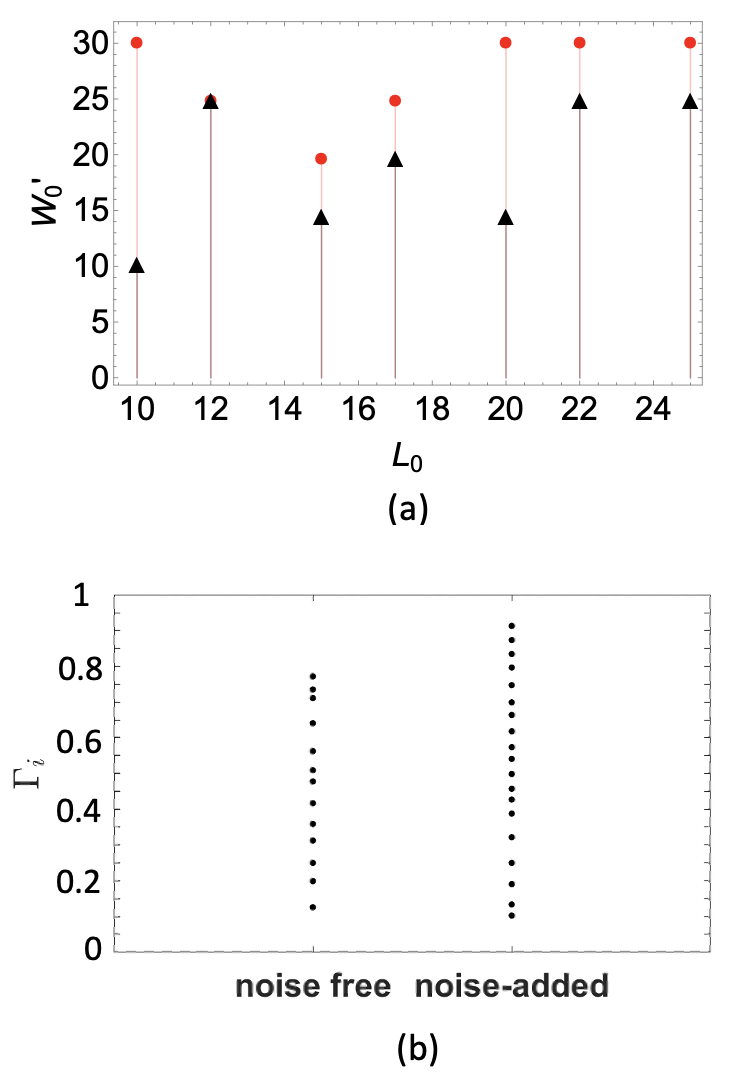}
  \caption{Effect of noise on the fracture of the crystalline sheet. (a) Plot of
  $W_{0}'$ versus the length of the crystalline sheet $L_0$ in the presence and
  absence of noise, as indicated by dots (red) and triangles (black),
  respectively. For $W\leq W_{0}'$, the interior of the crystalline sheet is
  overall free of defects in the fracture process (except a few transient
  defect events). (b) Comparison of the critical values $\Gamma_i$ in the
  intermittent plastic shear deformations in the presence and absence of noise.
  $L_0= 20.0$ and $W_0=19.9$.
	}
	\label{noise}
\end{figure}

\begin{figure*}[th] 
	\centering
	\includegraphics[scale=0.23]{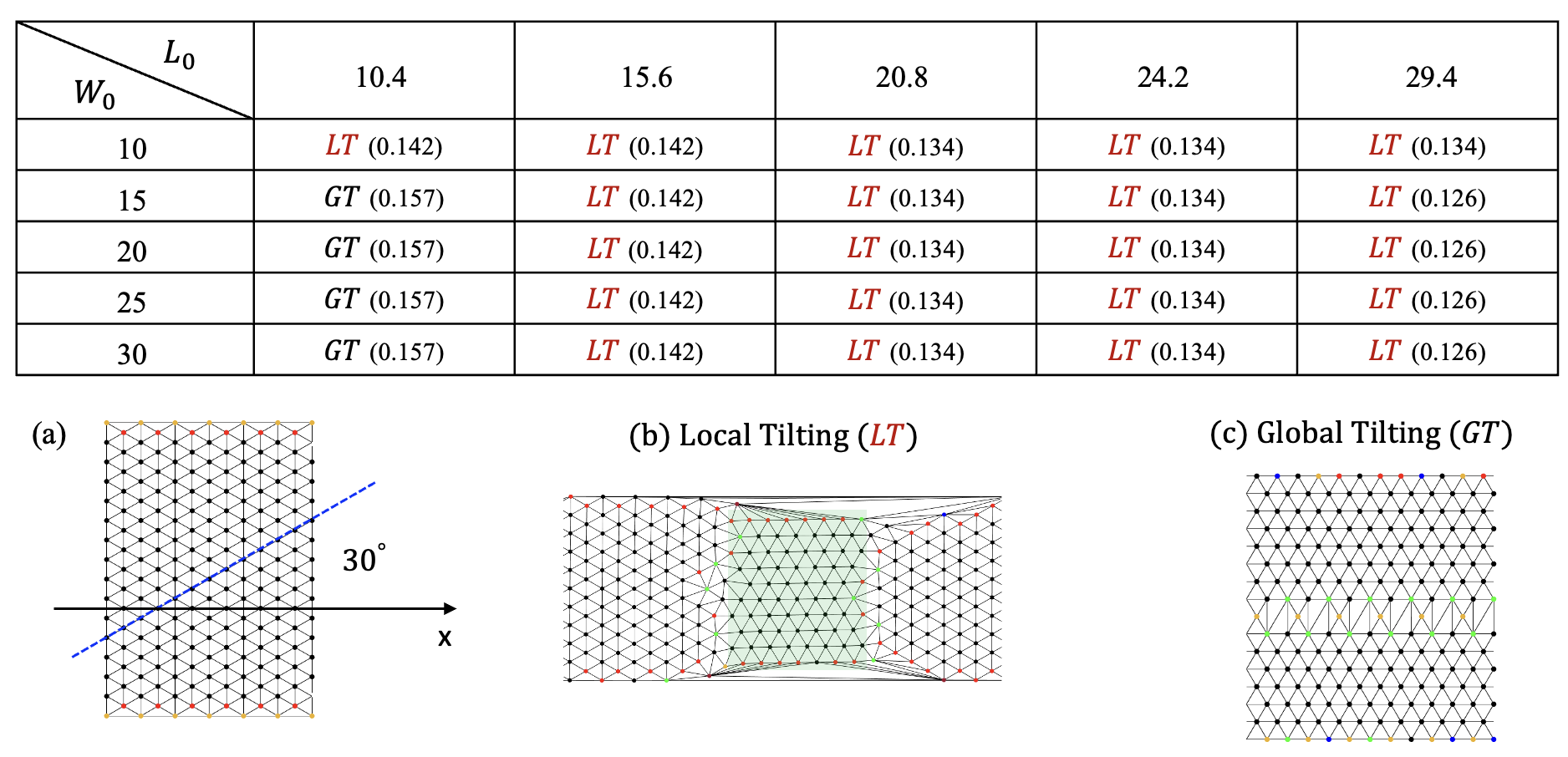}
  \caption{The abrupt tilting transition of the crystalline lattices initially at the
  tilt angle of $30$ degrees. The table shows the two distinct modes of tilting
  transition in crystalline sheets of varying geometry; the critical values of
  $\Gamma$ at which the tilting transition occurs are listed besides. (a) The
  undeformed crystalline lattice at the tilt angle of $30$ degrees. (b) In the
  local tilting (denoted as $LT$), the crystalline lattice is partly tilted to
  the $60$ degrees configuration; the tilted domain is highlighted in green. (c)
  In the global tilting (denoted as $GT$) as shown in (c), the entire lattice is tilted from the
  original $30$ degrees configuration to the $60$ degrees configuration. The
  subsequent fracture of the lattice upon further expansion is characterized
  by a series of intermittent plastic shear deformations, and it is similar to
  the case of the crystalline sheet whose tilt angle is $60$ degrees. 
	}
	\label{fig:new_orientation}
\end{figure*}

\subsection{Effects of noise and initial lattice orientation}

In this subsection, we discuss the effects of noise and initial lattice orientation on
the plastic deformation of the stretched crystalline sheet.

The stable states in the phase diagram in Table~\ref{phase_diagram} are obtained
by mechanical relaxations at zero temperature. We further investigate the
effect of noise on the fracture behavior of the crystalline lattice. To add noise in the
mechanical relaxation process, we introduce an uncertainty in the update of the
particle positions in simulations. Specifically, the new position of particle
$i$ that is originally at position $\vec{r}_i$ is located at $\vec{r}_i''=
\vec{r}_i' + c_0 \vec{\xi}$. $\vec{r}_i' = \vec{r}_i+\hat{F}_i s$ is the
new position of particle $i$ under the mechanical relaxation algorithm at zero
temperature, where $\hat{F}_i$ is the normalized force on particle $i$.  The
second term $c_0 \vec{\xi}$ represents a random displacement of maximum
magnitude $c_0$. $c_0=0.1s$, where $s$ is the step size.  $\vec{\xi}$ is a
random 2D vector whose orientation and magnitude are uniformly distributed in
the ranges of $[0, 2\pi)$ and $[0,1]$, respectively.  With the introduction of
the noise in the relaxation process, we still observe the intermittent plastic
shear deformations and the transition from the defect-free to the defect-based
plastic deformation as the width $W_0$ of the crystalline lattice of given
length $L_0$ is increased.

Simulations show that adding the noise at the level of $c_0=0.1s$ tends
to facilitate the glide motion of the dislocations and thus increase the
critical value of $W_0$ above which the defect-free to defect-based transition
occurs. In Fig.~\ref{noise}(a), we present the plot of $W_{0}'$ versus the
length of the crystalline sheet $L_0$ in the presence and absence of noise, as
indicated by dots (red) and triangles (black), respectively. For $W\leq W_{0}'$,
the interior of the crystalline sheet is overall free of defects in the fracture
process except a few transient defect events. These transient defect events in
the interior of the crystalline sheet do not affect the fracture process in
the sense that no stress-focusing effect around the defect is observed. For
example, in the system of $L_0=12$ and $W_0=19.9$ in the absence of noise, a
topologically neutral compound defect (consisting of one nine-fold disclination
surrounded by three five-fold disclinations) appears at $\Gamma_{19}=0.14$,
transforms into a vacancy (a pair of dislocations) at $\Gamma_{21}=0.16$, and
disappears at $\Gamma_{29}=0.22$.  In the system of $L_0=22$ and $W_0=10.4$, we
observe the appearance of a pair of isolated dislocations at $\Gamma_{35}=0.28$,
which disappears upon a gentle further expansion at $\Gamma_{37}=0.29$. From
Fig.~\ref{noise}(a), we see that the value of $W_{0}'$ is overall increased with
the introduction of the noise (red dots).  Here, we shall report an exceptional
case in the presence of noise at $L_0=15$. For $W\leq 19.9$, the fracture
process is defect-free. While the defect-free to defect-based transition occurs
at $W_0=25.1$, the fracture process becomes defect-free again at $W_0=30.3$.
This case implies the complication brought by the noise.

We also examine the effect of noise on the critical values $\Gamma_i$ in the
intermittent plastic shear deformations. A typical case is presented in
Fig.~\ref{noise}(b) for $L_0= 20.0$ and $W_0=19.9$. Comparison of the critical
values $\Gamma_i$ in the presence and absence of noise shows that the first
critical value $\Gamma_1$ is subject to a much smaller variation in comparison
with that of the critical value $\Gamma_f$, at which the crystalline lattice is
completely fractured.  The enhanced sensitivity of the highly stretched
crystalline sheet implies the appearance of many nearly-equivalent paths over
the energy landscape in the highly-stretched regime. This observation is
consistent with the variation of the critical values $\Gamma_i$ by increasing
the step size, which effectively introduces noise in the relaxation process.

In preceding discussions, we focus on the plastic deformations of crystalline
lattices at the tilt angle of $60$ degrees, i.e., the angle of the inclined
principal axis of the lattice with respect to the $x$ axis is $60$ degrees, as
shown in Fig.\ref{model}(a). In general, a seamless triangular lattice wrapping
the cylinder can be generated by a periodicity vector, which is characterized by
the phyllotactic index $[\ell, m, n]$, where $m\geq n$ and
$\ell=m+n$~\cite{mughal2012dense,mughal2014theory}. The lattice at the tilt
angle of $60$ degrees is represented by $[h, h, 0]$, where $h r_0$ is the
perimeter of the cylinder and $r_0$ is the lattice spacing.  Here, we consider
another typical case that the crystalline lattice wraps the cylindrical surface
at the tilt angle of $30$ degrees~\cite{mughal2012dense,mughal2014theory}; see
Fig.~\ref{fig:new_orientation}(a). The corresponding phyllotactic index is $[p,
p/2, p/2]$, where $p r_0$ is length of the spiral (i.e., the inclined
principal axis at the tilted angle of $30$ degrees wrapping the cylinder) within
a pitch.

Upon the gradual expansion of the cylinder at the same rate as in the case of
the tilt angle of $60$ degrees, it is found that the crystalline sheet exhibits
an abrupt transition from the $30$ degrees configuration to the $60$ degrees
configuration, either locally or globally depending on the geometry of the
crystalline sheet. The transition occurs abruptly under a gentle expansion of 
the cylinder by $0.7\%$. The results are summarized in the table in
Fig.~\ref{fig:new_orientation}. In the local tilting (denoted as $LT$), the
crystalline lattice is partly tilted to the $60$ degrees configuration as
shown in Fig.~\ref{fig:new_orientation}(b). The tilted domain is highlighted in
green.  Grain boundaries are formed at the interface of the domains of distinct
tilt angles. In the global tilting (denoted as $GT$) as shown in
Fig.~\ref{fig:new_orientation}(c), the entire lattice is tilted from the
original $30$ degrees configuration to the $60$ degrees configuration except the
thin horizontal belt; in some cases, the belt vanishes upon a further gentle
expansion of the cylinder. With the further expansion of the cylinder, the
subsequent fracture of the lattice is characterized by a series of intermittent
plastic shear deformations, and it is similar to the case of the crystalline
sheet whose tilt angle is $60$ degrees. The critical values of $\Gamma$ at which
the abrupt tilting transition occurs are recorded in the table in
Fig.~\ref{fig:new_orientation}.

\section{Conclusion}

In summary, we study the adaptations of crystalline sheets to uniaxial
stretching deformation and reveal the intermittent plastic shear deformations
leading to the complete fracture (disconnection) of the crystalline sheet
wrapping the cylinder. Systematic investigations of crystalline sheets of
varying geometry show that the fracture processes can be classified into the
defect-free and defective categories depending on the emergence of topological
defects. The computational approach reveals the characteristic mechanical and
geometric patterns arising in the crystalline sheet system in response to the
stretching deformation, including the shear-driven intermittent lattice tilting,
the vortex structure in the displacement field, and the emergence of mobile and
anchored dislocations as plastic excitations. Uniaxial stretching represents a
fundamental mechanical agitation and it also occurs in locally stretched
packings of particles in the contexts like crystal growth and 2D assembly. As
such, the results presented in this work may yield insights into the subtle role
of stretching deformation in triggering structural instabilities of 2D regular
packings of particles in general.

\section{Acknowledgements}

This work was supported by the National Natural Science Foundation of China
(Grants No. BC4190050).


\end{document}